# Magnetocaloric effect in some magnetic materials in alternating magnetic fields up to 22 Hz

A.M. Aliev[1*], A.B. Batdalov[1], L.N. Khanov[1], V.V. Koledov[2],
V.G. Shavrov[2], I.S. Tereshina[3], S.V. Taskaev[4]

[1]*Amirkhanov Institute of Physics of Daghestan Scientific Center, RAS, 367003 Makhachkala,
Russia*
[2]*Kotelnikov Institute of Radio Engineering and Electronics, RAS, Moscow 125009,
Russia*
[3]*Baikov Institute of Metallurgy and Materials Science RAS, Moscow, Russia*
[4]*Chelyabinsk State University, 454001 Chelyabinsk, Russia*

\* Electronic mail: lowtemp@mail.ru

Direct measurements of the magnetocaloric effect (MCE) in different materials (Gd, $Fe_{48}Rh_{52}$, $Ni_{43}Mn_{37.9}In_{12.1}Co_7$ and $Ni_{2.07}Co_{0.09}Mn_{0.84}Ga$) in alternating magnetic fields with frequencies $f \leq 22$ Hz and an amplitude $\Delta H = 6.2$ kOe are carried out. The MCE in Gd shows inconsiderable changes with field frequency. Near paramagnetic-ferromagnetic phase transition in $Ni_{43}Mn_{37.9}In_{12.1}Co_7$ Heusler alloy a slight reduction of MCE with frequency is observed. In weak alternating fields in materials with AFM-FM magneto-structural phase transitions ($Fe_{48}Rh_{52}$, $Ni_{43}Mn_{37.9}In_{12.1}Co_7$) it is not possible to get a structural contribution to overall MCE because of irreversibility of the transitions in these fields. Near magneto-structural phase transitions the MCE in these alloys has only magnetic contribution, and does not show a significant dependence on the magnetic field frequency. In $Ni_{2.07}Co_{0.09}Mn_{0.84}Ga$ Heusler alloy the MCE vanishes at frequencies about 20 Hz. The obtained results show the increase of frequencies of operating cycles is one of the powerful methods to improve the efficiency of magnetic refrigerators in case of Gd as a refrigerant.

1. Introduction

There are many problems towards the creation of magnetic refrigeration machines, which have a number of significant advantages over conventional cooling systems [1-6]. For the production of magnetic refrigerators materials with giant values of the magnetocaloric effect (MCE) and compact sources of strong magnetic fields are needed. Significant progress has been made recently in the production of such sources [7-8]. Various alloys with MCE reaching significant values have been synthesized [9-15]. Nevertheless, gadolinium and its alloys are still the most frequently used refrigerant [5-6].

For solid state refrigerators created on the basis of existing magnetocaloric materials to have no less efficiency than conventional gas-liquid devices it is necessary to increase their power density. One of the most effective ways to enhance cooling power is increase the frequencies of working cycles of refrigerating machines. Almost all currently existing prototypes of refrigerators operate at cycle frequencies up to 4 Hz [5-6], and only some of them operate at higher frequencies up to 8 and 10 Hz [16, 17]. Estimating the maximum adiabatic temperature change that can be achieved can never exceed 18 K/T, the more realistic upper limit lying somewhere in high single figures [18]. So to improve an efficiency of magnetic cooling machines on the basis of currently known giant MCE materials it is needed to increase the frequency of working cycles.

The efficiency of the refrigerating machine can only be improved by the increase of working cycle frequencies provided that magnetocaloric properties of the material do not change significantly with frequency of alternating (AC) magnetic field. In recent years dynamic methods of MCE study were developed [19-21] and many interesting results were obtained [22-29]. Nevertheless, some of questions concerning the operation of magnetic refrigerators at frequencies of tens or hundreds of Hertz remain open. There are at least two obvious problems that can occur in AC fields. First of all, it is probable dependence of the MCE value on the frequency of field. Due to the various relaxation processes the increase of the frequency of AC field can lead to a decrease of MCE value. In the Ref. [26] upper bound to the operational frequency in MCE based refrigerators are calculated without proper consideration of the magnetic relaxation. Even if we assume that the magnetization/demagnetization of the material takes place simultaneously with the switch on/off of the magnetic field, some time is required to get change in entropy of the lattice through the spin-lattice interaction. Another problem is the reduction of the ability of the working material to exchange energy with heat exchanger when the frequency of cycle increases. To improve the heat exchange processes it is necessary to increase the surface area of the working material of the refrigerator that can be achieved by making the latter in the form of thin plates [22-24].

In AC fields a degradation of magnetic properties of materials may occur. This is especially relevant for the materials with first order magnetostructural phase transitions, which are currently regarded as the most promising materials for magnetic refrigeration technology. In addition, the destructive effect of AC may consist in change of $T_C$ (with corresponding shift of the maximum of the MCE) and even physical destruction of the material. Currently, these questions have been properly addressed neither experimentally nor theoretically.

Investigation of the effect of strong AC magnetic fields at high frequencies on magnetic, properties of materials is of fundamental interest. From a practical standpoint, the studies are important to determine the upper frequency limit of use of them as refrigerants for different magnetocaloric materials.

Conventional MCE research methods are not suitable for study of magnetocaloric properties in AC magnetic field, since both indirect (estimation of MCE from the magnetization or heat capacity data) and direct ones are stationary techniques, i.e., either constant applied magnetic field is used in these experiments (to measure the magnetization or heat capacity) or the magnetic field switch-on technique (for direct measurements of MCE) [30].

## 2. Experiment

To measure the MCE in AC magnetic field a new technique has been used [31]. This technique has been tested on many materials giving results in good agreement with literature data, but at significantly higher temperature sensitivity. Bulk samples of the Heusler alloys [32], manganites [33], as well as other structures of magnetocaloric materials such as microwires [34], ribbon samples [35] and sandwich structures [36] have been studied. In contrast to other dynamic methods [19-21], the proposed one allows to study of MCE at frequencies up to 50 Hz.

Conventional electromagnets are not suitable for generating AC fields with frequencies higher than 1 Hz because of the coil inductance. Adjustable sources of dc magnetic field also do not generate variable fields with frequencies more than a few Hertz. To study the MCE in the fields with frequency up to 22 Hz, we have created a special source of alternating magnetic field comprising of four permanent magnets, each with an induction 0.62 T, fixed on the disk (Figure 1a). By rotating the disk with magnets via stepper motor it is possible to obtain AC magnetic fields with frequency up to 50 Hz. The form of the resulting field is shown in Figure 1b.

During measurements, it is necessary to take into account the amplification factor of the transformer preamplifier since at low frequencies it strongly depends on the frequency and the input impedance of the signal. Amplification factor was determined by measuring the AC voltage of known value, with a shape like the used AC magnetic field and with the input impedance equal to the resistance of the thermocouple. For this a signal generator AKIP-3407 was used.

With increasing frequency of the magnetic field an influence of the thermal inertia of the thermocouple also grows. To measure the MCE chromel-constantan thermocouple is used, which consists of junction of wires flattened to a thickness of 3-5 µm. In order to test the effect of inertia, the emf of the thermocouple was measured at frequencies up to 40 Hz. For that we used a miniature film heater through which an alternating current of different frequencies passes. A thermocouple was glued directly to the heater. With increasing the frequency of current, the magnitude of the temperature oscillations induced by heater reduces corresponding number of times, due to the reduction of heat generated per cycle. Slight deviation from linear decrease of signal is observed at frequencies up to 30 Hz, and at 30 Hz the deviation from the expected value of the thermocouple reaches about 12%. Significant influence of inertia begins to appear at frequencies above 30 Hz.

Sample sizes were almost identical, about 2.5x2.5x0.4 mm$^3$. The sample's plane *ab* and the magnetic field were parallel to each other. The AC magnetic field induces eddy currents, which can cause uncontrolled heating of the sample and the sample chamber. Eddy current does not affect on measured MCE value. It is seen on the figures given below, on which you can see that the effect occurs only near the phase transition temperatures, while eddy currents are induced in all studied temperature range. But eddy current can affect on the MCE value due to a magnetic field shielding effects. It is obvious that at frequencies of the experiment the eddy current will not affect on the measurement because the penetration depth of the AC magnetic field is more than the thickness of the studied samples. In addition, the thermocouple is glued to the surface of the sample, and a role of the core of the sample can be neglected. The geometry of the experiment also leads to minimizing the impact of eddy currents. Identical geometry of the experiment and sample sizes also means approximately equal values of a demagnetization factor for all samples, which is important when comparing the results for the different samples [37]. The influence of eddy currents in our experiments lies in a noticeable heating of the temperature insert with the sample at frequencies above 10 Hz. To control the temperature of the sample at high frequencies, it is necessary enhance heat dissipation by reducing the vacuum in the outer of thermal insert. For high frequency MCE studies the chamber is prefer to fabricate from high resistance material to avoid of heating of that due to eddy currents.

Frequency dependences of the MCE were measured in several different materials, including Gd, $Fe_{48}Rh_{52}$ alloy and $Ni_{43}Mn_{37.9}In_{12.1}Co_7$ and $Ni_{2.07}C_{0.09}Mn_{0.84}Ga$ Heusler alloys. Materials for the initial studies were selected based on the following criteria. The Gd is well studied magnetocaloric material with large value of the direct MCE in the vicinity of the second order magnetic phase transition [30]. The Gd is used in the majority of prototypes of magnetic refrigeration machines created to date [5-6]. $Fe_{48}Rh_{52}$ alloy is an example of a material with giant inverse magnetocaloric effect near antiferromagnetic-ferromagnetic (AFM-FM) first order phase transition [38, 39]. This transition is accompanied by sharp volume change. In $Ni_{43}Mn_{37.9}In_{12.1}Co_7$ Heusler alloys both direct and inverse MCE at Curie point $T_C = 430$ K and at magneto-structural ferromagnetic austenite - antiferromagnetic martensite phase transition are observed respectively (corresponding temperatures of start and finish of martensite and austenite transitions are $M_s = 285$ K, $M_f = 275$ K, $A_s = 304$ K and $A_f = 321$ K) [40]. In $Ni_{2.07}Co_{0.09}Mn_{0.84}Ga$ Heusler alloy the martensitic-austenite transition at $T_m=306$ K and the FM-PM magnetic transition at $T_C = 367$ K occur [41].

## 3. Results

The results of measurement of the MCE in Gd versus frequency of the field are shown in Figure 2. The maximum effect is observed at a temperature of about 294 K. It is obvious that with increasing frequency of AC magnetic field the magnitude of effect is slightly reduced. At 20 Hz the MCE is less pronounced than for 2 Hz by about 10%. There is a just slight shift of the maximum temperature of the MCE with frequency to lower temperatures. This effect can be explained by the increase in the disordering influence of AC field.

Giant inverse MCE is observed in the $Fe_{48}Rh_{52}$ alloy (Figure 3, inset). Despite the large value of MCE $\Delta T_{ad} = -7.5$ K and $-8.7$ K in a relatively small field $\Delta H = 18$ kOe in the heating and cooling protocols respectively, disadvantage of the alloy is narrow width of the effect at moderate magnetic field. At $\Delta H=18$ kOe width of half maximum does not exceed 6 K. In comparison with the value of MCE in moderate field, at $\Delta H = 6.2$ kOe MCE has considerably smaller magnitude ($\Delta T_{ad} = -0.27$ K). Small value of the MCE in weak fields at metamagnetic phase transition can be explained by the fact that MCE has two contributions, magnetic and structural. The application of a weak AC field to the sample in the antiferromagnetic state away from the phase transition point does not induce change in the lattice parameters. Near the phase transition the application of a weak field can induce transformation to a ferromagnetic state with the change in the lattice parameters due to its expansion, but when the field is switched off, inverse transition with the alteration of the lattice parameters does not occur due to the thermal hysteresis. Reversible transition in the $Fe_{48}Rh_{52}$ alloy with the change of the magnetic state and the lattice parameters can be observed only at high AC magnetic fields. The small value of the measured MCE in this alloy is only due to the magnetic contribution. Measurements of MCE show negligible frequency dependence in magnetic fields up to 20 Hz, and at 20 Hz the reduction of the effect is 10% approximately, as in the case of gadolinium. Similar observed frequency dependence of the MCE indicates that it can be explained by the thermal inertia of the thermocouple. In fact, we can say that in this frequency range magnetic contribution to the MCE does not depend on the field frequency. Stronger AC field are required to study the frequency dependence of the structural contribution to the MCE.

Heusler alloys also are considered as promising materials for use in magnetic refrigeration technology due to the large MCE at magnetostructural phase transition from the high temperature austenitic ferromagnetic phase to the low-temperature martensitic antiferromagnetic phase. Figure 4 shows that the behavior of frequency dependence of the MCE in the $Ni_{43}Mn_{37.9}In_{12.1}Co_7$ alloy near magneto-structural and FM-PM transitions differs substantially. MCE near the magneto-structural transition is small and does not depend on the frequency of field. At the same time, the MCE value near the FM-PM transition is large for such small change of the field, but there is significant dependence on the frequency of field (the reduction of the effect is more than 20% with frequency increase). Furthermore, with frequency increase a temperature of the maximum effect noticeably shifts toward lower temperatures. The behavior of the MCE in AC magnetic fields near magneto-structural transition in this composition may be similar to the that of $Fe_{48}Rh_{52}$ alloy i.e. the field amplitude is not sufficient to induce a reversible structural transition.

The surprising results are obtained in the study of the frequency dependency of the MCE in $Ni_{2.07}Co_{0.09}Mn_{0.84}Ga$ Heusler alloy (Fig. 5). The value of MCE at field change of 1.8 T with 0.3 Hz frequency is in agreement with [40] direct conventional measurements (see Inset in Fig. 5). In weak fields the MCE in the martensitic phase is not observed, although at $\Delta H=18$ kOe nonzero effect is observed down to liquid nitrogen temperatures. Strong frequency dependency of the MCE is observed in wide temperature range above and below the Curie point. At 20 Hz the MCE vanishes completely. It is hard to explain the strong frequency dependence of MCE in $Ni_{2.07}Co_{0.09}Mn_{0.84}Ga$ by relaxation of the magnetization/demagnetization processes because these times are usually of nanoseconds order. Strong frequency dependency of the MCE can be observed at magnetostructural transitions due to probable long term relaxation of structural transformation. Magnetostructural reversible AFM-FM transitions in the studied $Fe_{48}Rh_{52}$ and $Ni_{43}Mn_{37.9}In_{12.1}Co_7$ are not observed due to the small amplitude of the applied AC magnetic field. Apparently, it can be assumed that the magnetic transition FM-PM in $Ni_{2.07}Co_{0.09}Mn_{0.84}Ga$ alloy has some peculiarities. Without further studies of the structural and magnetic properties of the $Ni_{2.07}Co_{0.09}Mn_{0.84}Ga$ alloy we can only make some assumptions about the origin of the strong frequency dependency of the MCE.

According to [42], there is also the beginning of a structural transition at the FM-PM transition in some Heusler alloys. In this case, unlike the magneto-structural martensite-austenite transition, even a relatively weak fields can induce a reversible structural change, i.e. MCE, measured in an AC field, consists of two contributions, magnetic and structural. If the assumption of the existence of magnetic and structural phase transitions at the Curie point is true, vanishing of

MCE at high frequencies can be due to large relaxation times of the structural transition. Because of magnetostriction, the change of magnetic state induces structure transformation. Therefore, at large relaxation times of structural transition, frequency increase results to decrease in both structural and magnetic contributions to the overall MCE because magnetic state will not change while a structural transformation does not occur. The scenario is different from that takes place at AFM-FM magnetostructural transition, where the magnetic field is not sufficient to induce reversible transitions. In the latter there is a small MCE caused only by a change in the magnetic state. In order to establish the true nature of the strong frequency dependency of the MCE in $Ni_{2.07}Co_{0.09}Mn_{0.84}Ga$ alloy more research is needed, in particular, the magnetostriction in AC magnetic fields.

## 4. Conclusion

The results obtained in this paper do not answer the question about the mechanism of the frequency dependence of the MCE. The reduction of MCE with increase of the AC field frequency can be explained by spin-spin and spin-lattice relaxation processes, hysteresis loss, and the heat transfer between the sample and the thermocouple. Obviously, elucidation of the mechanism for the different behavior of the MCE in AC fields requires further research.

From our study the following conclusions can be made. In the region of second order ferromagnetic-paramagnetic phase transitions (such as in Gd as an example), there is no appreciable frequency dependence of the magnetocaloric effect in fields up to 22 Hz. Apparent slight change of magnetocaloric effect partially can be explained by thermal inertia of the thermocouple. Giant magnetocaloric effect in the materials with magneto-structural phase transitions is observed only under fields that induce a structural transition, whereby we have the magnetic and structural contributions to the resultant magnetocaloric effect. In the studied materials with magneto- structural AFM-FM transitions in weak AC magnetic fields is not possible to get a contribution from the structural transition. In $Ni_{2.07}Co_{0.09}Mn_{0.84}Ga$ alloy in the region of FM-PM transition there is well pronounced frequency dependence of the MCE, which may be a consequence of slow relaxation processes in some structural transformations. Thus, one of the requirements to the magnetocaloric materials "great value of MCE" should be replaced by the requirement of "great value of MCE in AC magnetic fields". These results indicate that for some magnetocaloric materials the upper frequencies of the AC in which they can be used as refrigerants are limited by a few hertz. It is necessary to carry out similar studies of other promising magnetocaloric materials.

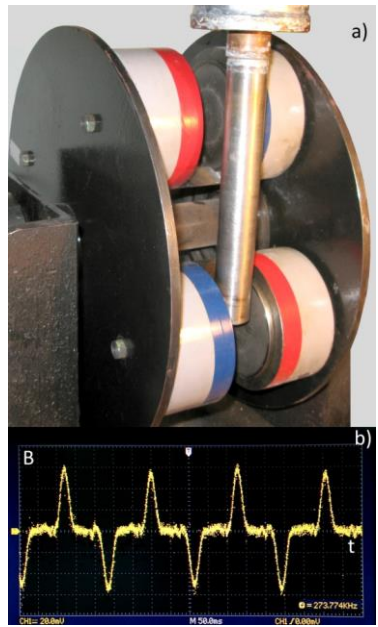

Figure 1. Photo of AC magnetic field source comprising of four permanent magnets, each with an induction B=0.62 T.

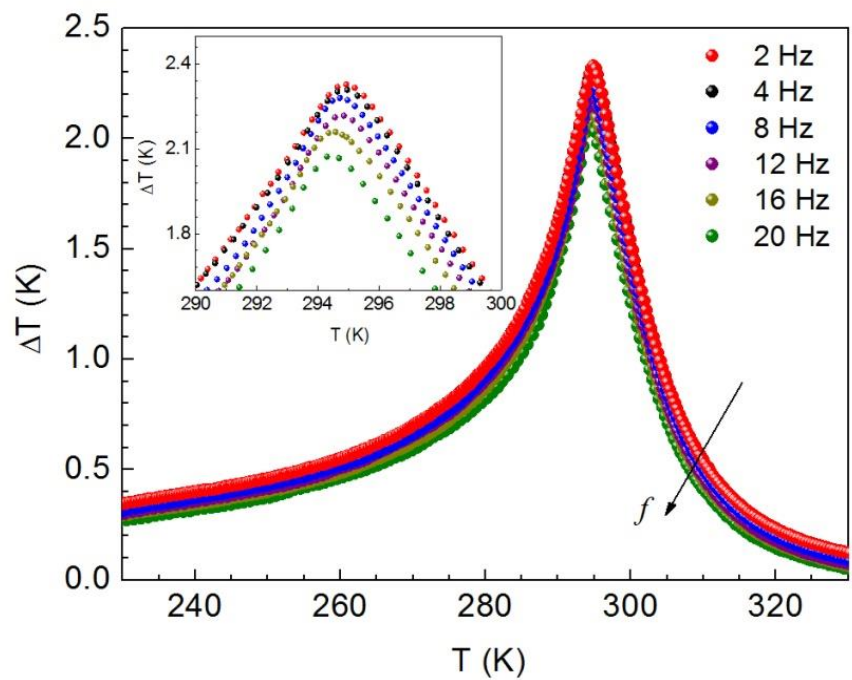

Figure 2. Temperature dependencies of MCE in Gd at different frequencies of AC magnetic field.

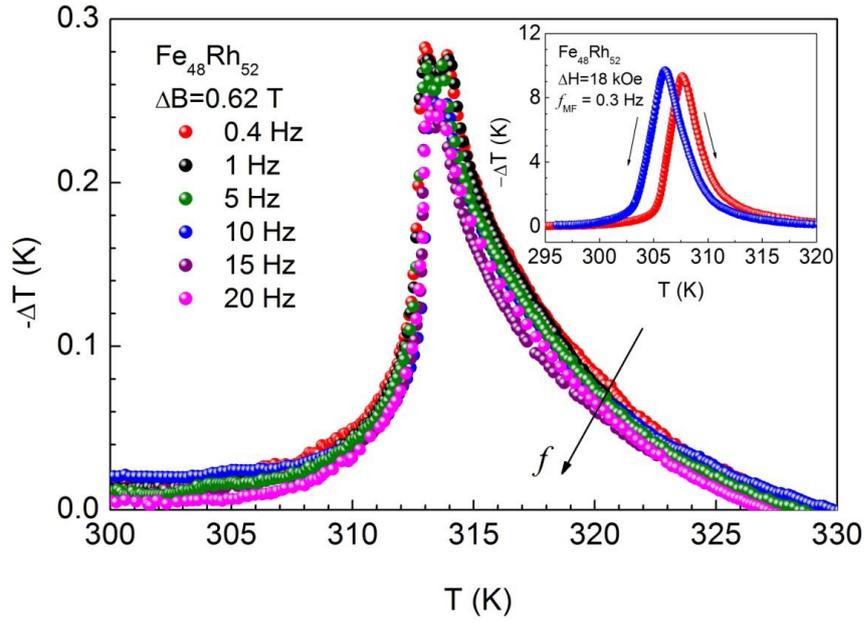

Figure 3. Temperature dependencies of MCE in Fe$_{48}$Rh$_{52}$ alloy at different frequencies of AC magnetic field. Inset – MCE in Fe$_{48}$Rh$_{52}$ at ΔH=18 kOe.

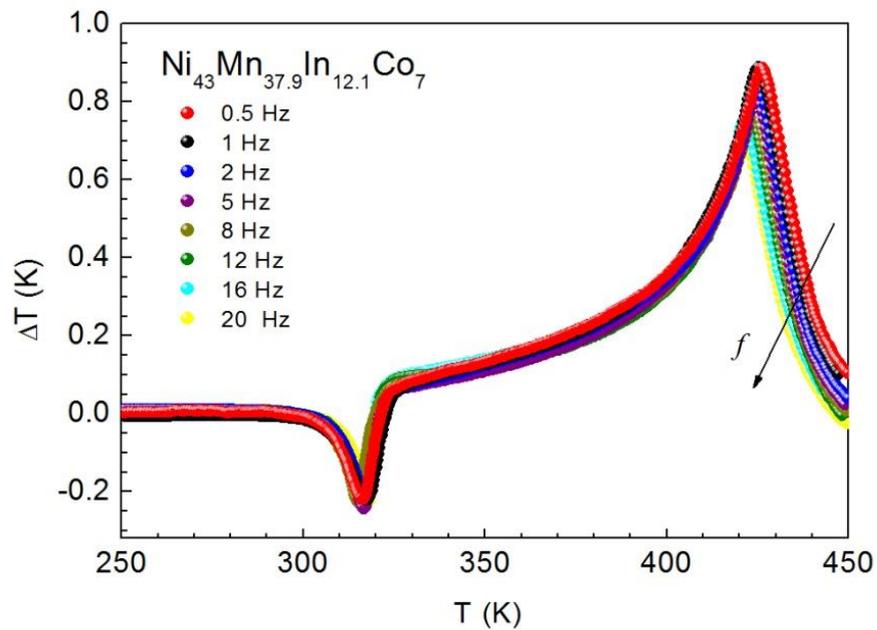

Figure 4. Temperature dependencies of MCE in Ni$_{43}$Mn$_{37.9}$In$_{12.1}$Co$_7$ Heusler alloy at different frequencies of AC magnetic field.

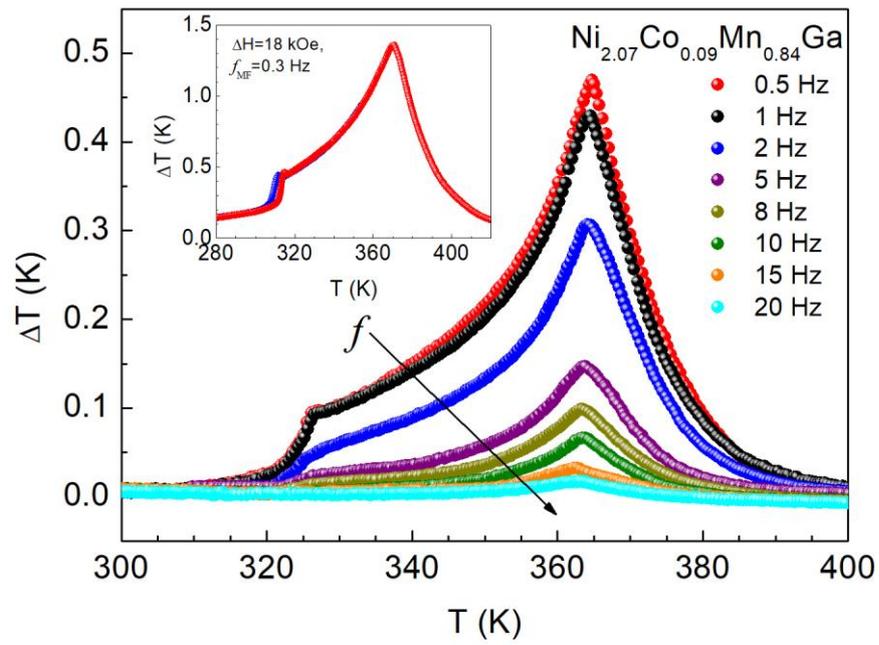

Figure 5. Temperature dependencies of MCE in Ni$_{2.07}$Co$_{0.09}$Mn$_{0.84}$Ga Heusler alloy at different frequencies of AC magnetic field. Inset – MCE in Ni$_{2.07}$Co$_{0.09}$Mn$_{0.84}$Ga at ΔH=18 kOe.